# Fundamental scaling laws of water window X-rays from free electron-driven van der Waals structures

Nikhil Pramanik[1*], Sunchao Huang[1*], Ruihuan Duan[2*], Qingwei Zhai[1], Michael Go[1], Chris Boothroyd[3,4], Zheng Liu[3] and Liang Jie Wong[1‡]

[1]School of Electrical and Electronic Engineering, Nanyang Technological University, 50 Nanyang Avenue, Singapore 639798, Singapore

[2]CINTRA CNRS/NTU/THALES, UMI 3288, Research Techno Plaza, Nanyang Technological University, 50 Nanyang Avenue, Singapore 637371, Singapore

[3]School of Materials Science and Engineering, Nanyang Technological University, 50 Nanyang Avenue, Singapore 639798, Singapore

[4]Facility for Analysis, Characterisation, Testing and Simulation (FACTS), Nanyang Technological University, 50 Nanyang Avenue, Singapore 639798, Singapore

‡Email: liangjie.wong@ntu.edu.sg

*These authors have contributed equally.

## Abstract

Water-window X-rays are crucial in medical and biological applications, enabling natural contrast imaging of biological cells in their near-native states without external staining. However, water-window X-ray sources whose output photon energy can be arbitrarily specified – a crucial feature in many high-contrast imaging applications – are still challenging to obtain except at large synchrotron facilities. Here, we present a solution to this challenge by demonstrating table-top, water-window X-ray generation from free electron-driven van der Waals materials, resulting in output photon energies that can be continuously tuned across the entire water window regime. In addition, we present a truly predictive theoretical framework that combines first-principles electromagnetism with Monte Carlo simulations to accurately predict the photon flux and brightness in absolute numbers. Using this framework, we theoretically obtain fundamental scaling laws for the tunable photon flux, showing good agreement with experimental results and providing a path to the design of powerful emitters based on free electron-driven quantum materials. We show that we can achieve photon fluxes needed for imaging and spectroscopy applications (over $10^8$ photons per second on sample) where compactness is important, and the ultrahigh fluxes of synchrotron sources are not needed. Importantly, our theory highlights the critical role played by the large mean free paths and interlayer atomic spacings unique to van der Waals structures, showing the latter's advantages over other materials in generating water window X-rays. Our results should pave the way to advanced techniques and new modalities in water-window X-ray generation and high-resolution biological imaging.

## Introduction

The water window regime is the part of the X-ray spectrum where water is transparent, but carbon is absorptive. In this regime, we may thus observe biological cells and its organelles in a non-destructive manner[1], as cells are composed of more than 70% water in which various carbon-based structures of interest are suspended. Continuously tunable water-window X-ray sources – referring to water-window X-ray sources whose photon energies can be arbitrarily specified – are highly sought after as they can be used to enhance the contrast of cellular images without the aid of staining agents. This enhanced contrast is obtained by leveraging the fact that different materials have different absorption spectra: as a result, images taken of the same sample at different photon energies will have differences that can be exploited to enhance image resolution. For example, imaging with photon energies below and above absorption edges can produce an elemental map with improved contrast for specific elements such as calcium and oxygen[2,3] – such elemental composition detection is vital for diagnosing diseases in patients.

Modern water-window microscopes rely on synchrotrons that produce continuously tunable X-rays, allowing the photon energy to be arbitrarily specified. However, the large size of synchrotrons limits their accessibility, fuelling the development of efficient table-top water-window microscopes[1,4–7]. The first of these lab-scale microscopes was the soft X-ray transmission microscope at the Royal Institute of Technology in Stockholm[8], which was based on a liquid-jet high-brightness laser-plasma source operating at fixed 2.58 nm wavelength. Other table-top water-window microscopes utilize electron-impact X-ray tubes[5–7,9] that emit characteristic X-rays, with fixed wavelengths, within the water window regime. Water window X-ray sources based on high harmonic generation[10–15] (HHG) enable the generation of coherent and ultrafast water-window X-ray pulses on a table-top scale, albeit with the requirement of high-intensity lasers.

Here, we introduce a versatile water-window X-ray source based on free electron-driven van der Waals (VdW) materials. We show that our source is highly complementary to existing water window X-ray sources, with distinct advantages especially when continuous energy tunability and compactness are important. It should also be noted that the water window regime, where our theories and experiments take place, constitutes a previously unexplored regime in light emission via free electron-driven vdW materials. Additionally, we theoretically obtain and experimentally reveal the fundamental scaling laws of the emitted photon flux. In the process, we present a truly predictive theoretical framework that combines first-principles electromagnetism and Monte Carlo simulations to accurately predict the output photon flux and brightness of free electron-driven quantum materials in absolute numbers. Our framework shows that the tunable photon flux increases linearly with current and sub-linearly with crystal thickness. We show that we can achieve fluxes of over $10^8$ photons per sec on sample, sufficient for a broad range of lab-based imaging applications. We demonstrate the ability to arbitrarily specify the photon energy peak by varying the tilt angle of the crystal and/or the driving electron energy, resulting in an output photon energy peak that can be continuously tuned across the water window regime. Our framework also reveals distinct advantages of using vdW materials as compared to conventional crystalline materials in our scheme: the larger interatomic 2D layer spacing possessed by vdW materials lead to larger mean free paths and electron interaction lengths, resulting in larger X-ray flux emitted in a narrower bandwidth. Our results should pave the way to compact, versatile water window X-ray sources for more discriminating medical and biological imaging.

## Results and Analysis

VdW materials have attracted tremendous attention due to their unique properties such as linear energy-momentum dispersion[16], massive intrinsic charge mobility[17,18], extreme electromagnetic confinement[19,20], quantum Hall effects[21–23], van Hove singularities[24], and tunable photon polaritons[25]. Among their many intriguing possibilities, vdW materials are excellent platforms for nanomaterial-based free-electron-driven X-ray sources[26–32] due to their high in-plane thermal conductivity[33] and strong plasmonic confinement[26]. Because of the large range of possible chemical combinations that allows control over the precise lattice constants defining the radiation spectrum, vdW materials and heterostructures[34–41] are tempting platforms for the creation of tunable X-rays[28,30,31,42], and for studying quantum effects in free-electron driven X-ray generation process[43,44].

Our study focuses on the generation of continuously tunable water-window X-rays using free-electron interactions with vdW materials, with the process depicted in Figure (Fig.) 1a. These materials consist of layered two-dimensional (2D) atomic planes, with larger interlayer spacing compared to conventional crystals, enabling us to produce high-quality monochromatic tunable water-window X-rays using a table-top setup. The X-rays are produced through the combination of two mechanisms, namely parametric X-ray radiation (PXR) and coherent bremsstrahlung (CB)[45–52]. These X-rays are tunable via three different mechanisms[30,31], 1) by varying the incident electron energy, 2) by varying the tilt angle of the crystal and 3) by varying the composition of the vdW material. The 2D atomic layers of the vdW materials acts like a nano grating which diffracts off the evanescent field of the incident electrons into propagating photons termed as parametric X-rays (PXR), whereas the other mechanism, coherent bremsstrahlung (CB), accounts for the X-ray emission by an undulated electron due to a series of periodic interactions with the crystal lattice. The resultant spectrum is the interference between PXR and CB. These two mechanisms in combined is referred to as parametric coherent bremsstrahlung (PCB) radiation. In our study, these two forms of X-ray radiation (PXR and CB) have the identical output X-ray photon energies[30,47] given by the equation

$$E = \hbar \frac{\boldsymbol{v} \cdot \boldsymbol{g}}{1 - (\boldsymbol{v} \cdot \hat{\mathbf{n}} \,/\mathrm{c})}, \tag{1}$$

where $\hbar$ is the reduced Planck constant, and $\boldsymbol{v}$ is the velocity of electron, $\boldsymbol{g}$ is the reciprocal lattice vector, $\hat{\mathbf{n}}$ is the unit vector along the observation direction, c is the speed of light in free space. We define the crystal tilt angle $\theta_{\mathrm{til}}$ as the angle between the [001] zone axis and the z axis as shown in Fig. 1a. The azimuthal rotation angle of the crystal with respect to the z-axis is denoted $\phi_{\mathrm{til}}$.

Combining this tunable water-window X-ray source with a range of X-ray optics can enable X-ray imaging of biological cells in a table-top setup. A schematic of our proposed X-ray imaging setup is shown in Fig. 1b. The X-rays emitted from the target material are selected by a monochromator which then directs a monochromatic X-ray beam towards the condenser optics. The condenser optics, for example an X-ray capillary tube or a zone plate, converges the incoming X-rays and focuses it on the desired sample surface after passing through an aperture. The X-rays transmitted through the sample is then detected by a X-ray CCD camera. An objective lens optics can also be used to focus the image onto the CCD camera. This setup can be easily altered to perform near-edge X-ray absorption fine structure (NEXAFS) spectroscopy by replacing the CCD camera with a spectrometer.

In Fig. 1b, the inset illustrates a potential application of the water window X-rays generated by this system. Using X-ray images captured above and below the calcium absorption edge, we can acquire high-contrast images of dense granules in blood platelet cells through a two-wavelength dichromography technique. The outcome of logarithmic subtraction[53] between the two X-ray images of the examined phantom blood platelet cell effectively highlights a significant contrast difference between the dense granules and other non-dense granules, facilitating accurate detection of dense granules (DG). This capability holds immense significance in the medical industry and proves invaluable for patient diagnosis[3]. This technique has shown promising results in accurately detecting DG from tunable water window X-rays generated in synchrotron facilities. Despite our source not possessing the ultrahigh fluxes of synchrotron-based sources, we show that we are able to obtain high-contrast images for DG detection (2000-fold enhancement), demonstrating the possibility to perform such diagnoses in a table-top setup. For further details on the blood platelet cell phantom model study using our X-ray source, please refer to the supplementary information (SI) section S8.

We experimentally and theoretically demonstrate that we can adjust the kinetic energy of incident electrons and tilt angle of the vdW crystal to finely tune the emitted X-ray photon energy within the water window regime. As shown in Fig. 1a, an electron beam impinges on a vdW material, resulting in the emission of water-window X-rays via PCB. The emitted X-rays are detected by an energy dispersive X-ray spectroscopy (EDS) detector. The results, shown in Fig. 1c and d, demonstrate a shift in peak intensity of the emission spectrum as incident electron energy is varied from 3.4 keV to 3.9 keV (Fig. 1c) and as tilt angle of the crystal $\theta_{\mathrm{til}}$ is adjusted in increments of 15°, 19°, 23° and 24° at a constant electron energy of 3.8 keV (Fig. 1d), with peak shift ranging from approximately 385 eV to 410 eV. The tilt angle of the target $\theta_{\mathrm{til}}$ is determined accurately by fitting the curves' peak intensity using Eq. (1). For more details on experimental data analysis please refer to SI section S10.

Figure 1e shows the regimes of table-top X-rays investigated using table-top electron sources[30,31,43,49,54–56], specifically focusing on electron energies below 1 MeV. Despite the abundance of literature on the relativistic regime[57–61], we restrict our attention in this work to table-top electron sources. Until now, expanding this domain to access water window X-rays had not been feasible. However, in this study, we have successfully demonstrated experimental evidence of accessing the water window X-rays. This was possible by exploiting the large interlayer $d$ spacing of vdW materials which allowed lower wavelength of photons for relatively higher kinetic energies of electrons when compared to conventional crystals.

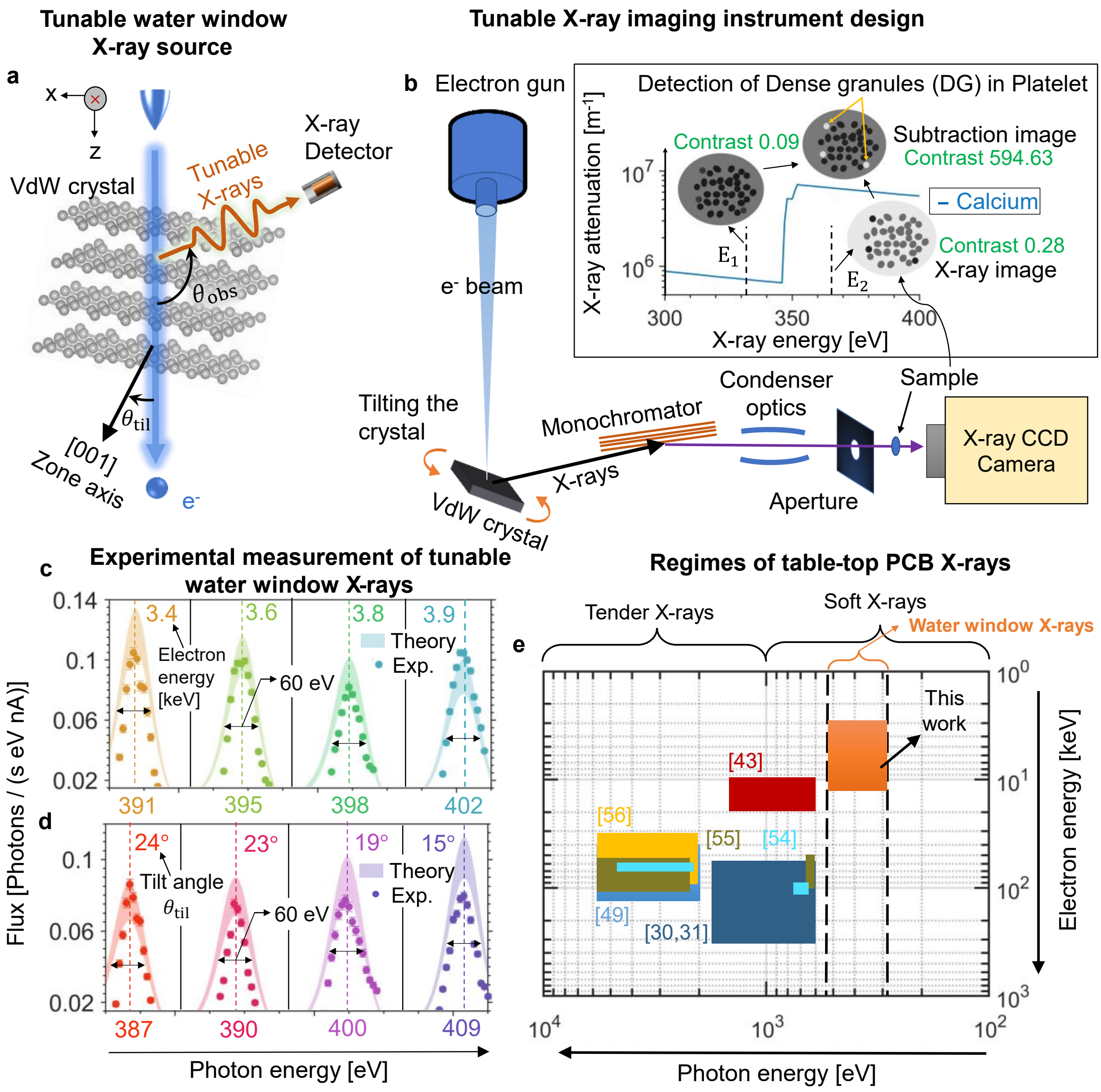

**Figure 1**. **Tunable water window X-rays from free electrons interacting with van der Waals (vdW) materials and X-ray imaging instrument design. a** shows the schematic diagram of an incident electron beam passing through a vdW material, generating water window X-rays via parametric X-ray radiation and coherent bremsstrahlung. The observation angle $\theta_{obs}$ is 119°. **b** shows the schematic of the table-top water window X-ray imaging setup. The inset illustrates a potential biological application demonstrating the accurate detection of dense granules in blood platelet cells using tunable water window X-rays with properties from our source. The Contrast for X-ray images is defined as flux difference between DG and other non-DG granules, while for the subtraction image, it denotes the difference between the logarithmic subtraction values of the flux from the two X-ray images. **c** demonstrates the photon energy spectrum tunability by varying incident electron energies in steps of 3.4 keV, 3.6 keV, 3.8 keV and 3.9 keV. The x-axis ticks are 50 eV apart with each graph partitioned by vertical black lines. **d** demonstrates the photon energy spectrum tunability by varying the tilt angle $\theta_{til}$ of the crystal by 15°, 19°, 23° and 24°. In panels c & d, the experimental data are represented by filled circles and theoretical predictions by shaded region. The vertical dotted lines indicate the peak photon energy predicted by Eq. (1). **e** illustrates the X-ray regimes investigated using table-top electron sources, with particular emphasis on the orange-colored region to highlight the novelty of this study in an unexplored regime.

In numerous X-ray imaging applications, X-ray photon flux is a critical factor. It is therefore important to appreciate the underlying laws governing the scaling of photon flux with various parameters. We have developed a theoretical framework combining first-principles electromagnetism with Monte Carlo simulations to accurately predict the tunable photon flux. Our theoretical framework showing the front-to-end simulation is described in Fig. 2b. The input parameters needed for the flux simulations are listed in Fig. 2a which consists of physical parameters related to incident electron beam, the target material and the monochromator. The arrows going from Fig. 2a to Fig. 2b shows which inputs are needed to simulate the quantities listed in Fig. 2b.

Our framework begins by simulating electron scattering inside the crystal via Monte Carlo simulations, here using the CASINO software[62]. The top right corner of the figure displays a Monte Carlo electron scattering simulation for a 50 keV electron beam interacting with a graphite target. The position of electron at each collision event is saved along with its kinetic energy. The positional information $\boldsymbol{r}$ and velocity $\boldsymbol{v}$ of each trajectory are subsequently inputted into the photon number spectral density formula $d^2N/d\omega d\Omega$ , which has been derived from Maxwell equations[31]. Considering the crystal's geometry and detector orientation, X-ray attenuation due to material absorption is included at every electron collision point using linear attenuation coefficient information from reference[63]. Next, the resultant intensity (after factoring in the attenuation) for each electron segment is summed over and finally averaged over total number of electrons to obtain the final spectrum per electron. The number of electrons in the scattering simulation is increased until convergence is achieved in the output spectrum. The emitted spectrum from the target is then combined with a X-ray geometric optics setup consisting of a monochromator and a condenser optics guided by an aperture to focus the X-rays on to the sample. Following Bragg's law, a narrowband X-ray beam is diffracted off the monochromator and focused on to the imaging sample using X-ray condenser optics.

Electron scattering suggests a decrease in intensity from the ideal case when there is no scattering. Therefore, using Monte Carlo electron scattering simulation we obtain an effective interaction length (or PCB interaction length), which can directly be used to estimate the emitted flux. This method allows for a faster numerical computation and is used to determine the flux given in Fig. 2e and 2f and to generate the theory plots in Fig. 3. We define the PCB interaction length as the electrons mean travel distance inside the crystal for which the deviation in the emitted photon energy is less than 1% of the peak photon energy, where the peak photon energy (given by Eq. (1)) is determined for the $\boldsymbol{g}$ corresponding to the most dominant reflecting plane.

Using our theoretical framework, we predict the scaling of the output photon flux through increasing the crystal thickness and incident electron current. This prediction is confirmed through experimental results presented in Fig. 2c and Fig. 2d. As shown in Fig. 2c, the radiation flux demonstrates a linear increase with an increase in the input current. The measurement error in flux is considered for fluctuations in the electron current, surface thickness variations at different locations (~5% uncertainty) and the error associated with the Poisson statistics of the photon counting process. Due to equipment limitations, we were only able to go up to 50 nA in current. However, extrapolating the current up to 10 mA would scale the output X-ray intensity to the order of $10^8$ photons per second as shown in the inset of Fig. 2c. Nevertheless, higher currents could result in a rise in target temperature, leading to melting or damage. To assess this risk, we performed thermal simulations for a graphite crystal exposed to an electron beam. The simulation took into account heat conduction inside the graphite crystal and the black body radiation mode of heat losses. The crystal was placed in a water-cooled copper holder, and the steady-state temperature at the graphite target center was well below its melting temperature for a 100 keV electron beam at 10 mA current with a 1 mm spot size. The detailed thermal analysis procedure with more test cases have been shown in the SI Section S4.

Increasing the thickness of the target crystal is another method to boost the X-ray intensity. Figure 2d depicts the increase in X-ray flux as the crystal's thickness rises from ~30 nm to ~170 nm. The thickness of the target crystal were measured in a Transmission electron microscope (TEM) using the convergent beam electron diffraction (CBED) technique[64–66] (SI Section S5). Combining our first-principles electromagnetism simulations with Monte Carlo simulations, we obtain the resulting flux emitted in the direction $\hat{\mathbf{n}}$, as

$$\frac{dN}{d\hat{\mathbf{n}}} \propto \left(\omega\left|\boldsymbol{v}\cdot\boldsymbol{E}_{\boldsymbol{g}s}\right|^2\right) i_c L^p \,, \tag{2}$$

where $N$ denotes the number of emitted photons, $\omega$ denotes the peak photon frequency, $\boldsymbol{v}$ is the velocity, $\boldsymbol{E}_{\boldsymbol{g}s}$ is the unit cell scattering amplitude[31,47], $i_c$ is the incident current and $L$ is the interaction length of electrons where $p \approx 3/4$ in our regime of study referencing sub-linear scaling of X-ray flux with increasing interaction length. Our numerical framework predicts our experimental results accurately, successfully capturing both the trend in flux variation and the precise flux values obtained from the experiment. In the context of electron scattering within vdW materials, this becomes particularly significant as it unveils an intriguing implication: the sublinear scaling of photon flux with electron interaction length.

Our proposed X-ray imaging setup utilizes a curved Si/W ellipsoidal mirror as the monochromator to enhance X-ray photon collection and diffraction towards the condenser optics. The ellipsoidal surface directs photons from the source to the second focus, where the condenser optics focus the X-rays into a narrow spot on the sample. Our front-to-end simulations using our theoretical framework (as described in detail in SI section S3) showed a flux in the order of $10^8$ photons per second on sample surface generated from a graphite crystal with just 30 keV electron beam as shown in Fig. 2e. Previous experiments have shown that such exposures are enough for imaging applications[4,13,67–69]. The corresponding flux spectrum for three different photon energies is shown in Fig. 2f, highlighting the narrow bandwidth of the radiation falling on the sample.

Notably, we can achieve fluxes in the order of $10^8$ photons per sec on sample, comparable to the existing lab-based water window X-ray sources [4–7,9–13,68,69]. Unlike these sources, our output photon energy peak is continuously tunable across the water window regime. Based on the manufacturers' specifications, we have considered an efficiency of 20% for the condenser optics (Wolter optics, zone plate or X-ray capillary tube) within the water window regime and the reflectivity spectrum of the multi-layer Si/W monochromator crystal is obtained from ref.[70]. Imaging at different wavelengths offers greater contrast imaging for different organic constituents within the biological cell because the X-ray transmission from different types of elements changes with the energy of the X-rays. The flexibility to tune the X-ray energy finely, unlike the existing table-top water-window X-ray imaging setups, allows our setup to participate in many applications requiring tunable X-rays. A narrow linewidth spectrum of water-window X-rays, which can be achieved with our setup, is particularly valuable for studying the constituents of biological cell organelles[71]. By restricting the illumination region of the ellipsoidal monochromator, we have control over the line shape of the spectrum that falls on the sample. This control is easily achieved by incorporating a movable beam stopper, as demonstrated in SI section S6.

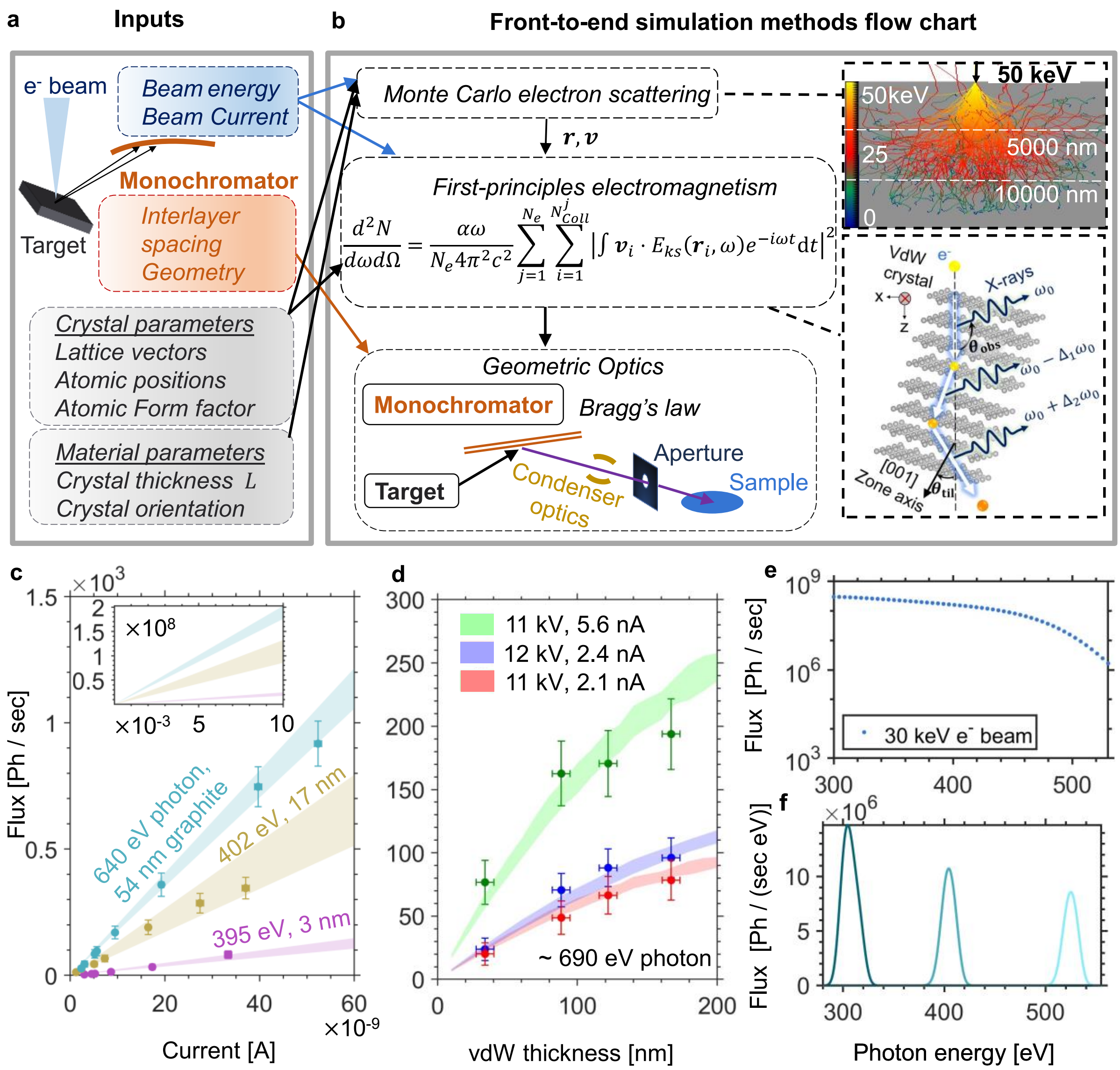

**Figure 2. Theoretical framework to simulate tunable photon flux in presence of electron scattering and experimental verification for flux scaling. a** Input parameters required for front-to-end (source-to-sample) simulation. **b** Front-to-end theoretical framework for simulation of flux falling on the sample. **c** Flux in photons per seconds (Ph/sec) generated by the source (graphite) varies linearly with the incident electron beam current, as detected by the energy dispersive X-ray spectroscopy (EDS) detector. The solid dots with error bars denote the experimental data while the shaded region refers to the theoretical flux simulated using the numerical framework for varying azimuthal crystal tilt angle $\phi_{\text{til}}$. The inset shows the scaled flux where the current is increased upto 10 mA. **d** shows the variation of the flux with increasing thickness of the crystal. **e** Simulated flux for various photon energies falling on the sample for 10 mA current using a 30 keV electron beam. **f** The spectrum of flux falling on sample for three different photon energies showing narrow bandwidth illumination. For a given electron energy the tilt angle $\theta_{\text{til}}$ of the crystal is adjusted (following Eq. (1)) to vary the emitted peak photon energy, and $\phi_{\text{til}} = 90°$.

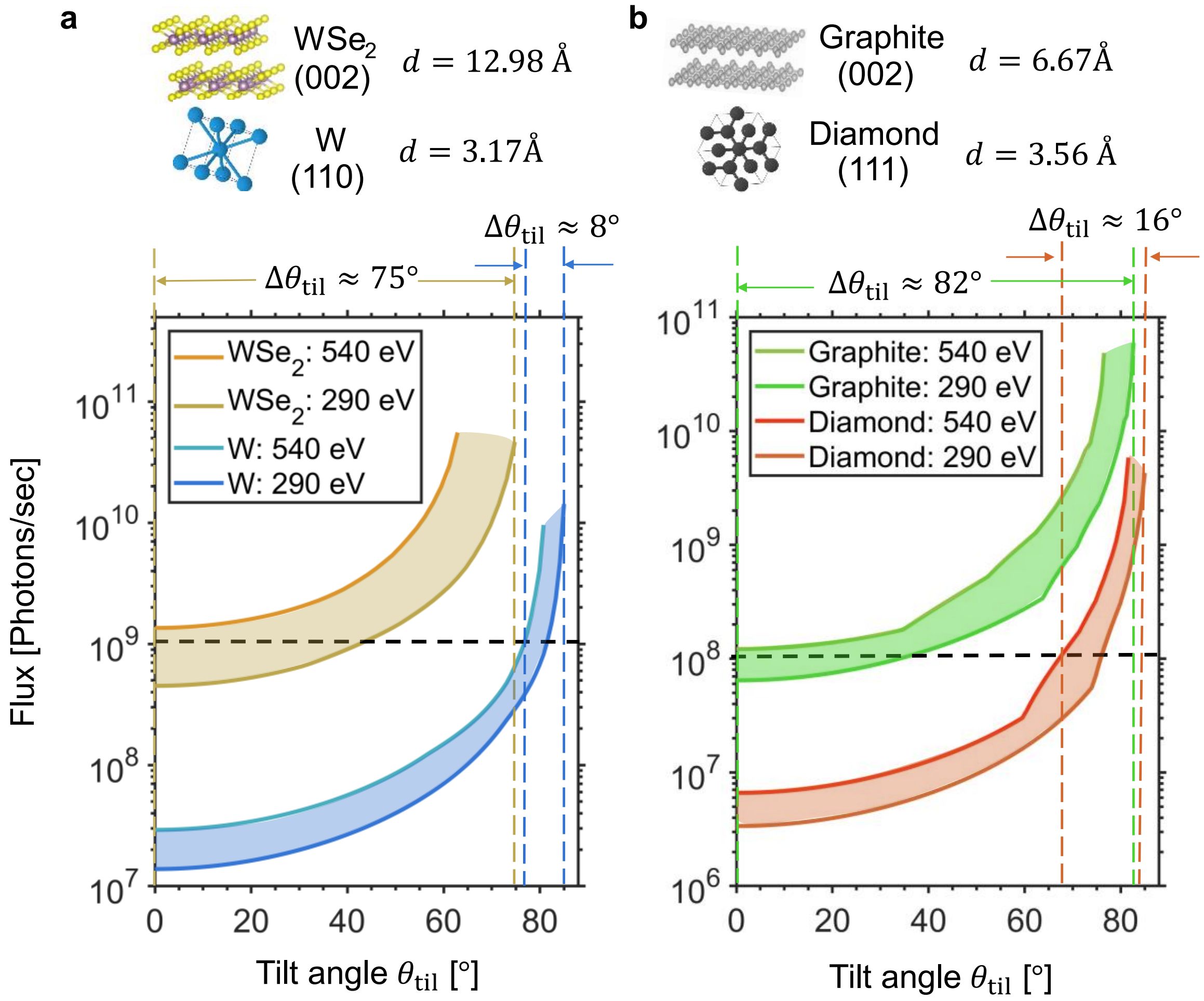

**Figure 3. Advantages of vdW crystals over conventional crystals in generating tunable water-window X-rays.** Crystals having higher structure factor, deeper electron interaction length, longer mean free path and higher $d$ spacing, tends to produce larger flux within a narrow spectral band. When compared to tungsten (W) ($\Delta\theta_{\mathrm{til}} \approx 8°$), $WSe_2$ generates much higher flux for a wide range of tilt angle ($\Delta\theta_{\mathrm{til}} \approx 75°$) as shown by the horizontal arrows in **a**. Similar is the case for diamond and graphite where diamond generates high flux within a very narrow tilt angle range $\Delta\theta_{\mathrm{til}}$ of just ~ 16 degrees whereas for graphite $\Delta\theta_{\mathrm{til}} \approx 82°$, as shown in **b**. This limited range makes it challenging to precisely tune the photon energy across the entire water window regime as the slightest mechanical errors can greatly affect the emitted photon energy. The incident electron energy along different curve is determined by Eq. (1) and is restricted to 1 MeV electron energy. Larger $d$ spacing for vdW materials leads to reduced density which leads to larger mean free path and longer interaction length. In addition, larger $d$ spacing requires incident electron with comparatively higher kinetic energy to generate a fixed energy photon, which in turn leads to even larger mean free path and interaction length leading to higher flux. Moreover, larger mean free path leads to narrower bandwidth and thus contributes to higher narrow-bandwidth flux for vdW materials. The y-axis in a and b corresponds to the flux integrated over 20 eV bandwidth, x-axis denotes the crystal tilt angle $\theta_{\mathrm{til}}$. The crystal is oriented such that the reciprocal lattice vector corresponding to the dominant reflecting plane aligns with z-axis. In this figure, $\phi_{\mathrm{til}} = 0°$, $\theta_{\mathrm{obs}} = 120°$, $\phi_{\mathrm{obs}} = 90°$, the electron current is 10 mA and the solid angle considered is 0.03 sr, respectively.

X-ray flux as a function of tilt angle of the crystal for vdW materials, $WSe_2$ and graphite, and conventional crystals, W and Diamond, are shown in Fig. 3. VdW materials offer an advantage over conventional crystalline materials because of their larger $d$ spacing (interlayer spacing), since larger $d$ spacing reduces the density and leads to comparatively larger mean free path and interaction length. The mean free path is defined as the electrons mean collision length inside the crystal for which the deviation in the emitted photon energy is less than 1% of the peak photon energy. Longer mean free path leads to a narrower bandwidth of emitted X-rays, increasing flux within this range. In addition, larger $d$ spacing allows comparatively higher electron energy (within table-top electron guns) for water window photon emission, which in turn leads to even longer mean free path and interaction length of the electron, enabling the production of higher flux within a narrow spectral range. A flow chart of this relation is shown in SI section S7. The advantage of longer $d$ spacing can be seen in a wider tilt angle range, with tungsten diselenide ($WSe_2$) producing higher flux compared to tungsten (W), as indicated by the range $\Delta\theta_{\text{til}}$ between the blue and brown vertical dashed line in Fig 3a.

Tungsten tends to produce a high flux of photons within a narrow tilt angle range $\Delta\theta_{\text{til}}$ of only approximately 8 degrees. Here, we have considered the range for higher flux region greater than $10^9$ photons/sec. This limited range of tilt angles presents a challenge in terms of precisely tuning the energy of the emitted photons across the entire water window regime. This is because even the slightest mechanical errors in the experiment can have a significant impact on the energy of the photons being produced, given the sensitivity of the system within such a narrow band of tilt angle variability. The same is the case for vdW material graphite and conventional material diamond, as shown in Fig. 3b, where the tilt angle range is considered for flux regime greater than $10^8$ photons/sec. Different material parameters used in Fig. 3 are given in SI section S7. To get the flux values in Fig. 3, we have integrated across 20 eV spectral band width of the X-ray peak emission. The X-ray peak spectrum was acquired from PCB radiation originating from the dominant reflecting plane with the highest structure factor, as indicated at the top of Fig. 3 next to each material. The electron energy used to generate the theoretical curves in Fig. 3 is determined using Eq. (1), with a limitation set at 1 MeV to suit the capabilities of table-top electron sources. As X-ray absorption within the material is dependent on geometric conditions, it has been ignored for the curves presented in Fig. 3. For instance, an electron beam focused close to the crystal's edge will result in minimal X-ray absorption.

## Discussion

VdW materials are an attractive platform for water-window X-ray sources based on free electron-driven quantum materials. The vdW X-ray source's frequency is dynamically tunable, scanning the wavelengths within the entire water window regime, unlike traditional X-ray tubes whose output X-ray peaks are fixed at the characteristic frequencies of the anode material. The proposed vdW material based free-electron-driven source is simpler to use and can be operated on a table-top setup. Furthermore, it requires neither high intensity lasers, as in high-harmonic generation[72] and laser plasma sources[73], nor highly relativistic electrons, as in undulator-based X-ray sources[74,75]. VdW based water-window continuously tunable X-ray sources thus can be a useful tool in X-ray microscopy and spectroscopy applications. Continuous photon energy tunability is also useful in near-edge X-ray absorption fine structure (NEXAFS) spectroscopy[10,11,76–80] including detection of elements with their absorption edges in the water window[71,81], which is the case for some essential elements forming organic matter (see supplementary information (SI) Table S1). Recent breakthroughs in shaping free electron wavepackets[82–93], laser driven electron acceleration[94] as well as X-ray generation into waveguide modes[95] could lead to greater control over and enhancement of the output radiation from our source.

A potential application of our continuously tunable, water-window X-ray source is the compact, accurate detection of calcium rich dense granule (DG) in blood platelet cells, helps in early detection of Osteoporosis disease. The standard method currently used to assess dense granule (DG) deficiency (DGD) in blood platelets is via transmission electron microscopy (TEM). Correct identification of DGs would be important to assess the status of platelets and DG-related diseases. However, due to electron-density based contrast mechanism in TEM, other granules such as $\alpha$-granules might cause false DG detection. Using X-ray transmission microscopy instead minimizes false DG detection of human platelets. DG can be identified by exploiting its high calcium content. Two X-ray transmission images are taken, one below the calcium (Ca) $L$-absorption edge and one above (post-edge region). A calcium map can be obtained from these two images revealing the regions of DG accurately. This method under scanning transmission X-ray microscopy (STXM) is successfully shown in synchrotron facilities using tunable water-window X-rays[3]. Although our source may not attain the ultrahigh photon fluxes of synchrotron sources, our studies show that the photon flux and other properties of our X-ray source are sufficient to generate high-contrast images for DG detection. This makes our source highly complementary to synchrotron sources: our source is even highly advantageous where compactness is important but the ultrahigh photon fluxes of synchrotrons are not needed. To directly validate the usefulness of our source in applications like biological imaging, we have studied phantom models of blood platelet cells in SI section S8, also presented in the inset of Fig. 1b (showing 2000-fold contrast enhancement). VdW based free-electron-driven sources provides a complementary tool to the present TEM based technique for conducting such studies for thin samples in a laboratory environment. Similar Calcium mapping in mineralized tissues by absorption difference imaging using two energy X-rays near the Calcium $L$-absorption edge, combining it with NEXAFS spectroscopy analysis provides a useful tool to quantify calcium concentration in tissues[2].

Other potential applications for our continuously tunable, narrowband water-window X-ray source includes NEXAFS spectroscopy for organic materials,[77,78] chemical mapping of biological samples[71], and various non-biological applications such as in microchemical analysis of inorganic materials: Paleo-botany[96]. NEXAFS, in particular, is used to investigate the structure of intermolecular bonds of polymers[77] by probing the electronic transition from the core level to the unoccupied states. Due of the distinct core binding energies that each element possesses, NEXAFS spectra can reveal elemental information. NEXAFS spectroscopy studies near absorption edges of various elements present in organic matter, such as an organic thin film, helps to determine the dynamic electronic properties[78]. Such applications can be performed by our proposed X-ray imaging setup by replacing the charge-coupled device (CCD) camera with a spectrometer.

Our results reveal that vdW materials have an undisputed edge over conventional materials due to the larger $d$ spacing of vdW materials, which provides greater interaction length and longer mean free path, resulting ultimately in higher flux within a narrower spectral bandwidth. The interaction length increases because of the reduction in the density of the crystal since the volume increases as the $d$ spacing of the crystal increases. The scaling of PCB intensity is a critical factor in many X-ray applications and understanding the fundamental laws governing this process is of utmost importance. Through this research, we have uncovered the fundamental scaling laws of PCB intensity, which show that it scales linearly with current and sub-linearly with thickness. We demonstrated these scaling laws through theoretical predictions and experimental validation, which has not been previously demonstrated. This discovery has significant implications for the field, as it opens up new possibilities for controlling and optimizing the intensity of PCB emissions from X-ray sources. Our findings have the potential to impact a wide range of industries that rely on X-ray technology, from medical imaging to material analysis and beyond.

In conclusion, we present a versatile, table-top X-ray source whereby continuously tunable water-window X-rays are generated from free electron-driven vdW structures. We present a truly predictive theoretical framework that combines first-principles electromagnetism and Monte Carlo simulations to accurately predict the output photon flux and brightness of free electron-driven quantum materials in absolute numbers. Using this framework, we theoretically obtain and experimentally reveal the fundamental scaling laws of the emitted photon flux. We show that the tunable photon flux increases linearly with current and sub-linearly with crystal thickness. Importantly, our theoretical framework also enabled the front-to-end simulation of water window imaging setups, enabling us to show the feasibility of fluxes exceeding $10^8$ photons per sec on sample. Such fluxes have been shown to be sufficient for a broad range of lab-based imaging and spectroscopy applications – but our setup also benefits from further enhancements in contrast made possible by the photon energy peak being continuously tuned across the water window regime. We demonstrate the ability to continuously vary the photon energy peak across the entire water window regime by varying the tilt angle of the crystal and/or the driving electron energy. Our framework also reveals the distinct advantages of vdW materials over conventional crystalline materials in our scheme, resulting in higher photon fluxes with narrower bandwidths. Our results bring a highly unique advantage of synchrotron-based water window X-ray sources – namely continuous photon energy tunability – to the table-top scale, and should unlock exciting prospects in high resolution imaging and spectroscopy for medical and biological applications.

## Methods

X-ray measurements: The X-ray measurements were performed in a FESEM JEOL7800. The emitted X-ray spectra were measured using a silicon drift energy dispersive X-ray spectroscopy (EDS) detector in the FESEM. The EDS detector was calibrated to measure the X-ray photon energies with an accuracy within ± 1 eV by measuring the Kα peaks of C, N, B, O, F & Si (see SI Section S9). The background radiation of the measured spectra was subtracted using NIST DTSA-II software[97,98]. The details of methods used to analyze the experimental data is described in SI Section S10.

Material synthesis and preparation: The graphite nanoflakes are exfoliated mechanically onto silicon substrates (with 285 nm $SiO_2$ film) and transferred to Au grids with the aid of the wet-transfer method. The Au grid is held by a copper TEM grid holder.

Theory and simulations for PCB intensity: See SI Section S2 and Section S3 for details.

Water-window X-ray imaging setup front-to-end simulations using our theoretical framework: See SI Section S3 for details.

Convergent beam electron diffraction (CBED) thickness measurements: See SI Section S5 for details.

**Data availability**

The data represented in Figs. 1-3 are available as Supplementary Information files. All other data that support the plots within this paper and other findings of this study are available from the corresponding author upon reasonable request. Source data are provided with this paper.

**Code availability**

All code that supports the plots within this paper are available from the corresponding authors upon reasonable request.